\documentclass[12pt]{article}

\usepackage[top=2.3cm, bottom=3.2cm, left=1.8cm, right=1.8cm]{geometry}
\usepackage{times,bbm,amssymb,amsmath,amsfonts,eucal}
\usepackage[shortlabels]{enumitem}
\usepackage{stmaryrd}
\usepackage{tikz}
\usetikzlibrary{calc}

\newtheorem{theorem}{Theorem}[section]

\newtheorem{definition}{Definition}[section]

\newcommand{\ck}{\text{$\boldsymbol{\mathsf K}$}}
\newcommand{\cs}{\text{$\boldsymbol{\mathsf S}$}}
\newcommand{\ci}{\text{$\boldsymbol{\mathsf I}$}}
\newcommand{\cj}{\text{$\boldsymbol{\mathsf J}$}}
\newcommand{\cm}{\text{$\boldsymbol{\mathsf M}$}}
\newcommand{\cb}{\text{$\boldsymbol{\mathsf B}$}}
\newcommand{\cl}{\text{$\boldsymbol{\mathsf L}$}}
\newcommand{\cx}{\text{$\boldsymbol{\mathsf x}$}}

\newcommand{\CL}{\text{$\boldsymbol{\mathsf {CL}}$}}
\newcommand{\dd}{\mathbb D}
\newcommand{\subs}{{\text{\sf sub}}}

\newcommand{\comp}{{\text{\sf comp}}}

\newcommand{\model}[1]{\big\llbracket #1 \big\rrbracket}

\usepackage{comment}

\begin{document}
\title{\bf On Universality of the $\cs$ Combinator}
\author{ Farrokh Vatan\footnote{NASA's Jet Propulsion Laboratory, California Institute of Technology, 4800 Oak Grove Drive, Pasadena, CA 91109. Email: Farrokh.Vatan@jpl.nasa.gov. 
\newline	
This work was done as a private venture and not in the author's capacity as an employee of the Jet Propulsion Laboratory, California Institute of Technology.}
}
\date{\today}
\maketitle
	
\begin{abstract}
In combinatory logic it is known that the set of two combinators $\cs$ and $\ck$ are 
universal; in the sense that any other combinator can be expressed in terms of
these two.  
We show that the $\ck$ combinator can not be expressed only in terms of the $\cs$ combinator. This will answer a question raised by Stephen Wolfram \cite{wolfram} as
``Is the $\cs$ combinator on its own computation universal?''
\end{abstract}

\section{Introduction}

Combinatory logic introduced by Sch\"{o}nfinkel \cite{schonfinkel} and developed
by Curry \cite{curry}. Wolfram's book, \cite{wolfram}, provides an extensive
historical background of its development. Here we consider combinatory logic as 
a rewiring (or substitution) system.

Here are the rewriting rules of some combinators, with the names given by Smullyan \cite{smullyan}:
\begin{align}
 \ck xy &\rhd x & (\text{Kestrel}), \label{K-equ}\\
 \cs xyz &\rhd xz(yz) & (\text{Starling}), \label{S-equ}\\
 \cb xyz &\rhd x(yz) & (\text{Bluebird}), \label{B-equ} \\
 \ci x & \rhd x  & (\text{Identity}), \label{I-equ} \\
 \cj xyzw & \rhd xy(xwz)  & (\text{Jay}), \label{J-equ} \\
 \cl xy &\rhd x(yy)  &(\text{Lark}),  \label{L-equ} 
\end{align}
\begin{align}
 \cm x   &\rhd  xx &(\text{Mockingbird}). \label{M-equ}
\end{align}
We denote the reflexive, transitive closure of $\rhd$ by $\rhd^\star$; i.e., 
$X\rhd^\star Y$ if and only if there is a sequence $X_1,\ldots, X_n$, $n\geq 1$, such that
$X_1=X$, $X_n=Y$, and $X_i\rhd X_{i+1}$, for $1\leq i\leq n-1$.

\vspace{4mm}
\begin{definition}[Terms of Combinatory Logic]
The language of combinatory logic consists of an infinite set of\/ {\em variables
$\cx_0, \cx_1,\ldots$} and two atomic constants\/ {\em $\ck$} and\/ {\em $\cs$},
called\/ {\em basic combinators}.
The set of expressions called\/ {\em combinatory logic terms}, or simply\/
{\em terms}, is defined inductively as follows:
\begin{enumerate}
\item
all variables and atomic constants are terms;
\item
if\/ $X$ and\/ $Y$ are terms, then so is\/ $(X\cdot Y)$. 
\end{enumerate}
A\/ {\em combinator} is a term having no occurrence of any variable. $\blacksquare$
\end{definition}

\vspace{4mm}
\noindent
In the following, for simplicity, we use
``$x$'', ``$y$'', ``$z$'', etc., 
to represent variables (distinct, unless otherwise stated). 
Also sometimes parentheses will be omitted following the convention of association to the left, 
so that $(((\cs x)y)z)$ will be abbreviated to $\cs x y z$, and 
$((\cs \ck)(\ck \cs))$ will be abbreviated to $\cs \ck (\ck \cs)$.
Also, we write $(X\cdot Y)$ simply as $(X Y)$ or $XY$.

According to the above definition, $\ck$ and $\cs$ are the only primitive
combinator and the other combinators defined by 
(\ref{B-equ})-(\ref{M-equ}) can be defined in terms of the two primitive ones; for example:
\begin{align*}
\cb & := \cs(\ck \cs)\ck,  \\
\ci & := \cs\ck\ck,    \\
\cl & :=  ((\cs((\cs(\ck\cs))\ck))(\ck((\cs((\cs\ck)\ck))((\cs\ck)\ck)))), \\
\cm & :=  \cs(\cs\ck\ck)(\cs\ck\ck) . 
\end{align*}

\vspace{4mm}
\begin{definition}
For a set\/ $\big\{C_1,\ldots,C_k\big\}$ of combinators,
$\CL(C_1,\ldots,C_k)$ is the set of the combinators built only from\/ $C_1,\ldots,C_k$ by means of application. $\blacksquare$
\label{CL_definition}
\end{definition}

\vspace{4mm}
\noindent
Thus $\CL(\ck,\cs)$ is the set of all combinators. There has been studies
of some subsets of $\CL(\ck,\cs)$.
Giraudo \cite{giraudo2} investigated $\CL(\cm)$ as an ordered set.
Sprenger and  Wymann-B\"{o}ni \cite{spencer} showed that $\CL(\cl)$ is
decidable. 
Probst and Studer \cite{probst} studied $\CL(\cj)$ to provide
an elementary proof of the strong normalization property of $\cj$.
Waldmann \cite{waldman} studied $\CL(\cs)$ to show that this
term rewriting system admits no ground loops. This extends
the known result of the absence of cycles. Also, the paper provides a
procedure that decides whether an $\cs$-term has a normal form.
In \cite{barendreg2017}, Barendregt et al. surveyed different problems regarding $\CL(\cs)$.

In this paper we investigate the universality of the combinator $\cs$. 
This is a natural question, as $\{\ck, \cs\}$ is  a universal basis for
combinators; in the sense that every rewriting rule can be represented
as a combinator in $\CL(\ck,\cs)$.
This is a question that Wolfram \cite{wolfram} has raised as 
``Is the $\cs$ combinator on its own computation universal?''
We provide a negative answer to this question: every combinator 
$\Sigma\in\CL(\ck,\cs)$ that satisfies the rewriting rule
$\Sigma x  \rhd^\star x$ does not belong to $\CL(\cs)$.

Our proof is based on a model of combinatory logic. We are not using the
elegant Scott's $D_\infty$ model (see, e.g., \cite{scott,hindley}), but a simpler 
set-theoretic model introduced by Engeler \cite{engeler1, engeler2,hall, hindley},
also mentioned by Plotkin \cite{plotkin}. We show that in this model every
combinator in $\CL(\cs)$ corresponds with a set that is 
``closed under a substitution rule.''
Then we show that interpretation of every combinator $\Sigma$ that satisfies the rewriting rule
$\Sigma x  \rhd^\star x$ does not has this property.
This proves that $\ck$ is not in  $\CL(\cs)$, in the sense that there is no
combinator $\Sigma_0\in\CL(\cs)$  such that $\Sigma_0xy \rhd^\star x$.

\section{A Model for Combinatory Logic}

Throughout this paper the notation ``$(a\rightarrowtail b)$'' means the ordered pair ``$(a,b)$'',
following the suggestion of \cite{engeler1}, ``to make notation mnemonic.''

\vspace{4mm}
\begin{definition} [The set $\cal G$]
We define the sets\/ $G_n$ recursively:
\begin{align}
G_0 &= \{0,1,2,\ldots\}; \label{G0-equ}\\
G_{n+1} &= G_n \cup \big\{(\alpha\rightarrowtail b) : \text{\em $\alpha\subseteq G_n$, 
    $\alpha$ is finite, and $b\in G_n$}\big\}. \label{Gn-equ}
\end{align}
Then
\[
{\cal G} = \bigcup_{n\geq 0} G_n.\quad \blacksquare
\]
\label{G-definition}
\end{definition}

\vspace{4mm}
\noindent
The members of $\cal G$ can be presented as trees.
In tree representation of $\alpha \rightarrowtail b$, the left branch is
labeled by the subset $\alpha\subseteq{\cal G}$ and the right branch by the 
element $b\in{\cal G}$ (see Figure \ref{fig-K}).

\vspace{4mm}
\noindent
We adopt the following definition of a model for the combinatory logic 
originally introduced by Engeler \cite{engeler1}, also Plotkin
\cite{plotkin} proposed a similar definition.

\vspace{4mm}
\begin{definition}[The Model $\dd$, \cite{engeler1,  engeler2, hall, hindley}]
The model\/ $\dd$ is consists of the background set 
\[ {\cal P} = \text{the set of all subsets of\/ ${{\cal G}}$}, \]
and the binary operation\/ $\bullet$ on\/ $\cal P$:
\[
M\bullet N = \big\{s : \text{\em there exists a finite $\alpha \subseteq N$ such that $(\alpha \rightarrowtail s)\in M$}\big\}.
\]
The interpretations of the basic combinators are defined as follows:
\begin{align}
\model{\ck} &= \big\{\big(\{t\}\rightarrowtail (\emptyset \rightarrowtail t)\big) : 
 t\in {{\cal G}} \big\}, 
       \label{K-model}\\
\model{\cs} &= \bigg\{ \bigg\{\tau\rightarrowtail \big(\{r_1,\dots,r_n\}\rightarrowtail s\big) \bigg\} \rightarrowtail 
 \bigg(\big\{\sigma_1\rightarrowtail r_1,\ldots,\sigma_n\rightarrowtail r_n\big\}
 \rightarrowtail\big(\sigma\rightarrowtail s\big)\bigg)  : \nonumber \\
&\hspace{18pt}  n\geq 0, r_1,\ldots,r_n\in{{\cal G}}, 
\sigma = \tau \cup (\cup_{i}\sigma_i)\in {\cal P},\text{\em $\sigma$ finite} \bigg\},  \label{S-model}   \\
\model{(X\cdot Y)} &= \model{X}\bullet \model{Y}.  \quad \blacksquare  \label{com-model}
\end{align}
\label{model-def}
\end{definition}

\begin{figure}
\begin{center}
\begin{tabular}{ccccccc}
\begin{tikzpicture}[scale=0.67]
	\tikzstyle{solid node}=[circle,draw,inner sep=2,fill=black]
	\node[solid node]{}
	child{node[solid node,label=left:{$\alpha$}]{}}
	child{node[solid node,label=right:{$b$}]{}}
	;
\end{tikzpicture}
&  & 
\begin{tikzpicture}[scale=0.67]
	\tikzstyle{solid node}=[circle,draw,inner sep=2,fill=black]
	\node[solid node]{}
	child{node[solid node,label=left:{$\{0\}$}]{}}
	child{node[solid node]{}
		child{node[solid node,label=left:{$\emptyset$}]{}}
		child{node[solid node,label=right:{$0$}]{}}}
	;
\end{tikzpicture} 
& & 
\begin{tikzpicture}[scale=0.67]
	\tikzstyle{solid node}=[circle,draw,inner sep=2,fill=black]
	\node[solid node]{}
	child{node[solid node,label=left:{$\emptyset$}]{}}
	child{node[solid node]{}
		child{node[solid node,label=left:{$\emptyset$}]{}}
		child{node[solid node]{}
			child{node[solid node,label=left:{$\{1\}$}]{}}
			child{node[solid node,label=right:{$0$}]{}}}
	}
	;
\end{tikzpicture} 
&  & 
\begin{tikzpicture}[scale=0.67]
	\tikzstyle{solid node}=[circle,draw,inner sep=2,fill=black]
	\tikzstyle{level 1}=[level distance=14mm,sibling distance=34mm]
	\tikzstyle{level 2}=[level distance=14mm,sibling distance=14mm]
	\node[solid node]{}
	child{node[solid node]{}
		child{node[solid node,label=left:{$\{0,1\}$}]{}}
		child{node[solid node]{}
			child{node[solid node,label=left:{$\emptyset$}]{}}
			child{node[solid node,label=right:{$1$}]{}}}}
	child{node[solid node]{}
		child{node[solid node,label=left:{$\{1\}$}]{}}
		child{node[solid node,label=right:{$2$}]{}}}
	;
\end{tikzpicture}	
\\
(a) & & (b) & & (c) & & (d)
\end{tabular}		
\end{center}
\caption{
(a) Tree representation of $(\alpha\rightarrowtail b)\in{\cal G}$, the left branch denotes a 
subset of $\cal G$ and the right branch a member of it;
(b) tree representation of $\{0\}\rightarrowtail (\emptyset\rightarrowtail 0)$;
(c) tree representation of $\emptyset \rightarrowtail \big(\emptyset\rightarrowtail (\{1\}\rightarrowtail 0)\big)$;
(d) tree representation of 
$\big\{\{0,1\}\rightarrowtail (\emptyset \rightarrowtail 1)\big\}\rightarrowtail (\{1\}\rightarrowtail 2)$.
}
\label{fig-K}
\end{figure}

\begin{figure}
\begin{center}
\begin{tikzpicture}
\tikzstyle{solid node}=[circle,draw,inner sep=2,fill=black]
\tikzstyle{level 1}=[level distance=14mm,sibling distance=70mm]
\tikzstyle{level 2}=[level distance=10mm,sibling distance=14mm]
\tikzstyle{level 3}=[level distance=10mm,sibling distance=14mm]
\node[solid node]{}
child{node[solid node]{}
child{node[solid node,label=left:{$\tau$}]{}}
child{node[solid node]{}
	child{node[solid node,label=left:{$\{r_1,\ldots,r_n\}$}]{}}
	child{node[solid node,label=right:{$s$}]{}}}}
child{node[solid node]{}
child{node[solid node,label=left:{$\{\sigma_1\rightarrowtail r_1,\ldots, \sigma_n\rightarrowtail r_n\}$}]{}}
child{node[solid node]{}
	child{node[solid node,label=left:{$\tau\cup(\cup_i\sigma_i)$}]{}}
	child{node[solid node,label=right:{$s$}]{}}}}
;
\end{tikzpicture}	
\end{center}
\caption{
Tree representation of a generic member (\ref{S-model}) of $\model{\cs}$; here $\tau,\sigma_i$ are finite subsets 
of ${\cal G}$ and $s, r_i$ are members of ${\cal G}$.
}
\label{fig-S}
\end{figure}

\vspace{4mm}
\noindent
In the original definition \cite{engeler1} of the model $\dd$, the interpretation of $\ck$ was
defined as 
\[
\big\{\big(\alpha\rightarrowtail (\beta \rightarrowtail t)\big) :
\alpha,\beta\subseteq {\cal G}, t\in\alpha,
\text{$\alpha$ and $\beta$ are finite} \big\}.
\]
Here we use the simpler definition of \cite{engeler2}.

Figure \ref{fig-K} (b) shows a tree representation of a member of $\model{\ck}$ and 
Figure \ref{fig-S} shows a tree representation of a generic member of $\model{\cs}$.

\vspace{4mm}
\begin{theorem}[\cite{engeler1,  hall, hindley}]
For subsets\/ $M$, $N$, and\/ $L$ of\/ ${\cal G}$, we have
\begin{align}
\model{\ck}\bullet M \bullet N & = M, \label{K-model-equ}\\
\model{\cs}\bullet M \bullet N \bullet L &= M\bullet L\bullet(N\bullet L). 
            \label{S-model-equ}
\end{align}
\end{theorem}

\vspace{4mm}
\noindent
{\bf Example 1.} The combinator $\ci$, defined by rewriting rule (\ref{I-equ}).
In $\CL(\ck,\cs)$ , the combinator $\ci$ is define as $\ci := \cs \ck \ck$,
because
\[
\ci  x =\cs \ck \ck x \rhd \ck x (\ck x) \rhd x.
\]
In fact, $\ci$ also can be defined as $\cs \ck C$, where $C\in\CL(\ck,\cs)$ is an arbitrary combinator. Then
\begin{align*}
\model{\ci} & = \model{\cs \ck \ck} \\
&= \model{\cs}\bullet \model{\ck}\bullet \model{\ck} \\
&= \big\{s: \exists \alpha_1, \alpha_2 \subseteq \model{\ck}
\text{ such that $\big(\alpha_1 \rightarrowtail (\alpha_2\rightarrowtail s)\big)\in\model{\cs}$}\big\} \\
&= \big\{s: \exists t\in {\cal G}, \exists\alpha_2 \subseteq \model{\ck}
\text{ such that $\left(\big\{\{t\}\rightarrowtail(\emptyset\rightarrowtail t)\big\} \rightarrowtail (\alpha_2\rightarrowtail s)\right)\in\model{\cs}$}\big\}.
\end{align*}
Comparing the last condition with (\ref{S-model}), it follows that here $n=0$, $\alpha_2=\emptyset$, 
and $s= \big(\{t\}\rightarrowtail t\big)$. Therefore,
\begin{equation}
\model{\ci} = \big\{\big(\{t\}\rightarrowtail t\big): t\in{\cal G} \big\}. 
\label{I-model}
\end{equation}
Note that if we used the definition $\ci = \cs \ck C$, for some other combinator
$C$, then we would get the same result. $\blacksquare$

\vspace{4mm}
\noindent
{\bf Example 2.}
Using (\ref{K-model}) and (\ref{I-model}), 
the interpretation of the combinator $\ck\ci$ is 
\begin{align*}
	\model{\ck\ci} & = \model{\ck}\bullet\model{\ci} \\
	&= \big\{X: \exists\alpha\subseteq\model{\ci}\text{ such that } 
	(\alpha\rightarrowtail X)\in\model{\ck}\big\} \\
	&= \big\{\big(\emptyset \rightarrowtail(\{t\}\rightarrowtail t)\big) : t\in{\cal G}\big\}.
\end{align*}
Let
\[
\ck^{**} := \ck(\ck\ci).
\]
Then
\[
\ck^{**} x y z = \ck(\ck\ci)xyz \rhd \ck\ci yz \rhd \ci z \rhd  z.
\]
The interpretation of the combinator $\ck^{**}$ is 
\begin{align*}
\model{\ck^{**}} &= \model{\ck(\ck\ci)} \nonumber\\
& = \model{\ck}\bullet\model{\ck\ci} \nonumber\\
&= \big\{X: \exists\alpha\subseteq\model{\ck\ci}\text{ such that } 
(\alpha\rightarrowtail X)\in\model{\ck}\big\} \nonumber\\
&= \big\{\big(\emptyset \rightarrowtail\big(\emptyset \rightarrowtail(\{t\}\rightarrowtail t)\big)\big) : t\in{\cal G}\big\}.
\quad\blacksquare
\end{align*}

\section{Substitution}

\subsection{Templates for the generic members}
The equations (\ref{K-model}), (\ref{S-model}), and (\ref{I-model}) define templates
for the generic member of \model{\ck}, \model{\cs}, and $\model{\ci}$, respectively. 
Each template consists of variables, like $\tau$ and $s$ in (\ref{S-model}), which
represent an arbitrarily finite subset or a member of $\cal  G$.

The same is true for any combinator $\Sigma\in\CL(\ck,\cs)$, in the sense that there is
a templates for the generic member of $\model{\Sigma}$, consists of variables denoting
either arbitrarily finite subsets or members of $\cal  G$. 
To obtain this template,
suppose that $\Sigma= \Sigma_1\cdot\Sigma_2$, where $\Sigma_1,\Sigma_2\in\CL(\ck,\cs)$. 
There are the templates 
${\cal T}_1 = Y\rightarrowtail X$ and ${\cal T}_2$ for $\Sigma_1$ and $\Sigma_2$, respectively. Consider ${\cal T}_2\rightarrowtail X$ and modify $X$ to $X'$
such that ${\cal T}_2\rightarrowtail X'$ follows the template ${\cal T}_1$.
Then $X'$ is the template for the generic member of $\model{\Sigma}$.

\vspace{4mm}
\noindent
{\bf Example 1.}
Consider the combinator $\Sigma_1=(\cs\cdot\ck)$. From (\ref{S-model}), 
the template for the generic member of $\model{\cs}$ is
\begin{equation}
{\cal T}_1 =  
\bigg\{\tau\rightarrowtail \big(\{r_1,\dots,r_n\}\rightarrowtail s\big) \bigg\} \rightarrowtail 
\bigg(\big\{\sigma_1\rightarrowtail r_1,\ldots,\sigma_n\rightarrowtail r_n\big\}
\rightarrowtail\big((\tau\cup\cup_i\sigma_i)\rightarrowtail s\big)\bigg),
\label{S-template}
\end{equation}
for $n\geq 0$ and $\sigma = \tau \cup (\cup_{i}\sigma_i)$. From(\ref{K-model}),
the template of $\model{\ck}$ is
\[
{\cal T}_2 = \big(\{t\}\rightarrowtail(\emptyset\rightarrowtail t)\big).
\]
Now ${\cal T}_2\rightarrowtail {\cal T}'$ follows the template ${\cal T}_1$ if
and only if
\begin{align*}
\big(\{t\}\rightarrowtail(\emptyset\rightarrowtail t)\big)  &= \big(\tau\rightarrowtail (\{r_1,\dots,r_n\}\rightarrowtail s)\big),\\
{\cal T}' &= \big(\big\{\sigma_1\rightarrowtail r_1,\ldots,\sigma_n\rightarrowtail r_n\big\}
\rightarrowtail\big((\tau\cup_i\sigma_i)\rightarrowtail s\big)\big).
\end{align*}
Thus, $\tau=\{t\}$, $n=0$, and $s=t$. Therefore,
the template for the generic member of $\model{\cs\cdot\ck}$ is
\[
{\cal T}' = \big(\emptyset \rightarrowtail (\{t\}\rightarrowtail t)\big).
\]
In another words,
\[
\model{\cs\cdot\ck} =\big\{\big(\emptyset \rightarrowtail (\{t\}\rightarrowtail t)\big): t\in{\cal G}\big\}.
\]

\vspace{4mm}
\noindent
{\bf Example 2.}
Consider the combinator $\Sigma_2=(\cs\cdot\cs)$, and the template 
${\cal T}_1$ of equation (\ref{S-template}) for the generic member of $\model{\cs}$.
Now, 
${\cal T}_1\rightarrowtail {\cal T}''$ follows the template 
of $\model{\cs}$ of the following form
\[
\bigg\{\tau'\rightarrowtail \big(\{r'_1,\dots,r'_n\}\rightarrowtail s'\big) \bigg\} \rightarrowtail 
\bigg(\big\{\sigma'_1\rightarrowtail r'_1,\ldots,\sigma'_n\rightarrowtail r'_n\big\}
\rightarrowtail\big((\tau'\cup\cup_i\sigma'_i)\rightarrowtail s'\big)\bigg)
\]
if and only if
\begin{align*}
{\cal T}_1 &= \big(\tau'\rightarrowtail (\{r'_1,\dots,r'_n\}\rightarrowtail s')\big),\\
{\cal T}'' &= \big(\big\{\sigma'_1\rightarrowtail r'_1,\ldots,\sigma'_n\rightarrowtail r'_n\big\}
\rightarrowtail\big((\tau'\cup\cup_i\sigma'_i)\rightarrowtail s'\big)\big).
\end{align*}
Thus, 
\begin{align*}
\tau' &= \big(\tau\rightarrowtail \big(\{r_1,\dots,r_n\}\rightarrowtail s\big) \big), \\
r'_i &= \big(\sigma_i\rightarrowtail r_i\big), \quad 1\leq i\leq n,\\
s' &= \big((\tau\cup\cup_i\sigma_i)\rightarrowtail s\big).
\end{align*}
Therefore, ${\cal T}''$, the template for the generic members of $\model{\cs\cdot\cs}$, has the following form
\[
{\cal T}''= \bigg\{\sigma'_1\rightarrowtail(\sigma_1\rightarrowtail r_1),\ldots,\sigma'_n\rightarrowtail(\sigma_n\rightarrowtail r_n)\bigg\}\rightarrowtail
\bigg(\big(\tau'\cup\cup_i \sigma'_i\big)\rightarrowtail\big((\tau\cup\cup_i\sigma_i)\rightarrowtail s\big)
\bigg).
\]

Thus ${\cal T}''$ is the template for the generic members of $\model{\cs\cdot\cs}$;
i.e.,
\begin{align*}
\model{\cs\cdot\cs} &= \bigg\{
\bigg\{\sigma'_1\rightarrowtail(\sigma_1\rightarrowtail r_1),\ldots,\sigma'_n\rightarrowtail(\sigma_n\rightarrowtail r_n)\bigg\}\rightarrowtail
\bigg(\big(\tau'\cup\cup_i \sigma'_i\big)\rightarrowtail\big((\tau\cup\cup_i\sigma_i)\rightarrowtail s\big)
\bigg): \\
&\hspace{30pt} n\geq 0, 
\tau'=\big(\tau\rightarrowtail \big(\{r_1,\dots,r_n\}\rightarrowtail s\big) \big), 
s\in{\cal G}, r_i\in{\cal G},\text{$\tau$, $\sigma_i$, and $\sigma'_i$ finite subsets of $\cal G$}\bigg\}.
\end{align*}

\begin{theorem}
If\/ ${\cal T}$ is the template for the generic member of\/ $\model{\Sigma}$, 
where\/ $\Sigma\in\CL(\cs)$, then\/ ${\cal T}$ does not contain 
$\{t\}$, as a variable denoting a subset of $\cal G$. 
\label{template-theorem}
\end{theorem}

\noindent
{\bf Proof.}
Note that variables of (\ref{S-template}), the template for the generic member of
$\model{\cs}$, denote either members of $\cal G$ or finite subsets of it; i.e.,
no variable of the form $\{t\}$ as a variable representing a subset.
We prove the theorem by induction on the number of occurrences of $\cs$ in $\Sigma$. Then the base case, where $\Sigma=\cs$, is obvious. 
For the induction step, suppose that $\Sigma = \Sigma_1 \cdot\Sigma_2$, where
$\Sigma_1, \Sigma_2\in\CL(\cs)$. Let ${\cal T}_1$ and ${\cal T}_2$ be
templates for the generic members of $\model{\Sigma_1}$ and $\model{\Sigma_2}$,
respectively. By induction hypothesis,  ${\cal T}_1$ and ${\cal T}_2$
do not contain any $\{t\}$, as a variable denoting a subset of $\cal G$.
The template $\cal T$ for the generic member of $\model{\Sigma}$ is obtained 
from ${\cal T}_1\rightarrowtail {\cal T}$ by forcing it to follows the template ${\cal T}_1$. This process does not introduce any subset variable of the form
$\{t\}$. $\blacksquare$

\subsection{Companion}

First, some useful definitions and notations.

\begin{definition} [$B_0$ and $B_\mu$]
Let
\begin{align*}
B_0  &= \big(\{0\}\rightarrowtail 0\big), \\
B_\mu  &= \big(\{0\}\rightarrowtail \mu\big). 
\end{align*}
The value of the integer $\mu\geq 1$ will be determined later. $\blacksquare$
\label{X01-definition}
\end{definition}

\vspace{4mm}
\begin{definition}[$B_0$-Base]
A member\/ $X\in{\cal G}$ has\/ {\em $B_0$-base} if and only if
\begin{equation}
X = \big(\alpha_1\rightarrowtail (\cdots (\alpha_n\rightarrowtail B_0)\cdots)\big),
\label{B0-base-equation}
\end{equation}
where\/ $n\geq 0$ and\/ $\alpha_i\subseteq {\cal G}$ is finite. In the special
case of\/ $n=0$,  $B_0$ has\/ $B_0$-base. 
$\blacksquare$
\label{base-definition}
\end{definition}

\vspace{4mm}
\begin{definition}[$B_\mu$-Substitution]
Suppose\/ $X\in{\cal G}$ has\/ $B_0$-base and is of the general form\/
{\em (\ref{B0-base-equation})}. 
The\/ {\em $B_\mu$-substitution of $X$}, denoted as\/ $\subs_\mu(X)$, is
\[
\subs_\mu(X) = \big(\alpha_1\rightarrowtail (\cdots (\alpha_n\rightarrowtail B_\mu)\cdots)\big). \quad\blacksquare
\]
\label{substitution-definitin}
\end{definition}

\vspace{4mm}
\begin{definition} [$B_\mu$-Companion]
Suppose that \/ $X\in\model{\Sigma}$, of the form\/ {\em (\ref{B0-base-equation})},
for\/ $\Sigma\in\CL(\cs)$, has\/ $B_0$-base. Let\/ ${\cal T}$ be the template for the
generic member of\/ $\model{\Sigma}$. Therefor, $X$ is obtained from\/ $\cal T$
by substituting variables of\/ $\cal T$ by members or finite subsets of\/ $\cal G$.
Then there are two possible cases. {\em (i)} There is variable\/ $s$ which is substituted by
a\/  $B_0$-base\/ $Y\in{\cal G}$ to obtain\/ $X$. {\em (ii)} There are variables\/
$t$ and\/ $\sigma$ which are substituted by\/ $0$ and\/ $\{0\}$, respectively,
to obtain\/ $X$. Then the\/ {\em $B_\mu$-companion of\/ $B_0$}, denoted by\/ $\comp_\mu(X)$, is
obtained as follows: in case\/ {\em (i)} by replacing the variable\/ $s$ by
$\subs_\mu(Y)$; in case\/ {\em (ii)} by replacing the variable\/ $t$ by\/ $\mu$.
Here we assume that the integer\/ $\mu$ is bigger than any number appearing in\/ $X$.
$\blacksquare$
\label{companion-definition}
\end{definition}

\vspace{4mm}
\begin{theorem}
If\/ $\Sigma\in\CL(\cs)$, $X\in\model{\Sigma}$, $X$ has\/ $B_0$-base,
and\/ $\mu$ is bigger than any number appearing in\/ $X$,
then\/ $\comp_\mu(X)\in\model{\Sigma}$.
\label{companion-theorem}
\end{theorem}

\noindent
{\bf Proof.}
If $\comp_\mu(X)$ is obtained using rule (i) of Definition \ref{companion-definition},
then obviously $\comp_\mu(X)\in\model{\Sigma}$. 
If the rule (ii) is used, then the theorem follows from Theorem \ref{template-theorem}.
$\blacksquare$

\section{Combinators generated by $\cs$}

There are combinators in $\CL(\ck,\cs)$ which define the same rewriting rule as $\ci:= \cs\ck\ck$.
For example, $\cs \ck (\cs  \ck\cs\ck) \rhd^\star \cs\ck\ck$.
Also, for the combinator
\[
\Sigma_0 = \cs(\cs (\cs (\cs \ck)  (\cs(\ck\ck)\cs (\ck\ck) \ci))) (\ck \ci))\ck, 
\]
we have $\Sigma_0 x\rhd^\star x$, for all $x$; while it is not the case that
$\Sigma_0\rhd^\star \cs\ck\ck$.
The following theorem shows that interpretation of such combinators in $\dd$
is a super set of $\model{\ci}$.

\vspace{4mm}
\begin{theorem}
Let\/ $\Sigma\in\CL(\ck, \cs)$ such that $\Sigma x \rhd^\star x$, for all $x$. 
Then $\model{\ci}\subseteq \model{\Sigma}$. 
\label{sigma_theorem}
\end{theorem}

\noindent
{\bf Proof.}
From (\ref{I-model}), it is enough to show that
$\big(\{t\}\rightarrowtail t\big)\in\model{\Sigma}$,
for every $t\in{\cal G}$.
From the relations (\ref{K-equ}), (\ref{S-equ}), and (\ref{com-model})-(\ref{S-model-equ}) it follows that for every $M\subseteq {\cal G}$,
\[
\model{\Sigma}\bullet M = M.
\] 
Let $t\in{\cal G}$. Then
\begin{align*}
\{t\} &= \model{\Sigma}\bullet \{t\}  \\
&= \big\{s: \exists\alpha\subseteq\{t\} 
\text{ such that $\big(\alpha\rightarrowtail s\big)\in\model{\Sigma}$} \big\} .
\end{align*}
If $\alpha=\emptyset$, then  
$\big(\emptyset\rightarrowtail  t\big)\in\model{\Sigma}$, which implies 
$t\in \emptyset=\model{\Sigma}\bullet \emptyset$.
Therefore, $\alpha=\{t\}$; and 
$\big(\{t\}\rightarrowtail t\big)\in\model{\Sigma}$.
$\blacksquare$

\vspace{4mm}
\begin{theorem}
Let\/ $\Sigma\in\CL(\cs)$. Then it is not the case that\/
$\Sigma x \rhd^\star x$, for all $x$. 
\end{theorem}

\noindent
{\bf Proof.}
Suppose, by contradiction, that $\Sigma x \rhd^\star x$, for all $x$.
From (\ref{I-model}) and Theorem \ref{sigma_theorem}, it follows that $B_0=\big(\{0\}\rightarrowtail 0\big)\in\model{\ci}\subseteq\model{\Sigma}$. 
Note that from proof of Theorem \ref{sigma_theorem}, 
$\model{\Sigma}\bullet M = M$, for every
$M\subseteq {\cal G}$.
Then from the Theorem \ref{companion-theorem}
it follows that $\comp_\mu(B_0)=\big(\{0\}\rightarrowtail \mu\big)\in\model{\Sigma}$;
which implies 
$\mu\in \{0\} =\model{\Sigma}\bullet \{0\}$.
This contradicts the assumption $\mu>0$.
$\blacksquare$

\end{document}